\documentclass[pra,aps,twocolumn,eqsecnum,superscriptaddress,showpacs,amsmath]{revtex4-1}

\usepackage{graphicx}
\usepackage{latexsym}

\newcommand{\vc}[1]{\boldsymbol{#1}}

\begin{document}

\title{       Phase diagram and spin correlations of the Kitaev-Heisenberg model: \\
              Importance of quantum effects }

\author{      Dorota Gotfryd }
\affiliation{ Institute of Theoretical Physics, Faculty of Physics,
              University of Warsaw, Pasteura 5, PL-02093 Warsaw, Poland }
\affiliation{ Marian Smoluchowski Institute of Physics, Jagiellonian University,
              prof. S. {\L}ojasiewicza 11, PL-30348 Krak\'ow, Poland }

\author{      Juraj Rusna\v{c}ko }
\affiliation{ Central European Institute of Technology,
              Masaryk University, Kamenice 753/5, CZ-62500 Brno,
	      Czech Republic }
\affiliation{ Department of Condensed Matter Physics, Faculty of Science,
              Masaryk University, Kotl\'a\v{r}sk\'a 2, CZ-61137 Brno,
              Czech Republic }

\author{      Krzysztof Wohlfeld }
\affiliation{ Institute of Theoretical Physics, Faculty of Physics,
              University of Warsaw, Pasteura 5, PL-02093 Warsaw, Poland }

\author{      George Jackeli}
\affiliation{ Institute for Functional Matter and Quantum Technologies,
              University of Stuttgart, Pfaffenwaldring 57, D-70569 Stuttgart, Germany}
\affiliation{ Max Planck Institute for Solid State Research,
              Heisenbergstrasse 1, D-70569 Stuttgart, Germany }

\author{      Ji\v{r}\'{\i} Chaloupka }
\affiliation{ Central European Institute of Technology,
              Masaryk University, Kamenice 753/5, CZ-62500 Brno,
	          Czech Republic }
\affiliation{ Department of Condensed Matter Physics, Faculty of Science,
              Masaryk University, Kotl\'a\v{r}sk\'a 2, CZ-61137 Brno,
              Czech Republic }

\author{      Andrzej M. Ole\'s }
\affiliation{ Marian Smoluchowski Institute of Physics, Jagiellonian University,
              prof. S. {\L}ojasiewicza 11, PL-30348 Krak\'ow, Poland }
\affiliation{ Max Planck Institute for Solid State Research,
              Heisenbergstrasse 1, D-70569 Stuttgart, Germany }

\date{\today}

\begin{abstract}
We explore the phase diagram of the Kitaev-Heisenberg model with nearest
neighbor interactions on the honeycomb lattice using the exact diagonalization
of finite systems combined with the cluster mean field approximation, and
supplemented by the insights from the linear spin-wave and second--order
perturbation theories. This study confirms that by varying the balance between
the Heisenberg and Kitaev term, frustrated exchange interactions stabilize
in this model four phases with magnetic long range order: N\'eel phase,
ferromagnetic phase, and two other phases with coexisting antiferromagnetic
and ferromagnetic bonds, zigzag and stripy phases.  They are separated by two
disordered quantum spin-liquid phases, and the one with ferromagnetic Kitaev
interactions has a substantially broader range of stability as the neighboring
competing ordered phases, ferromagnetic and stripy, have very weak quantum
fluctuations.  Focusing on the quantum spin-liquid phases, we study spatial
spin correlations and dynamic spin structure factor of the model by the exact
diagonalization technique, and discuss the evolution of gapped low-energy spin
response across the quantum phase transitions between
the disordered spin liquid and magnetic phases with long range order.
\end{abstract}


\maketitle

\section{Introduction}

Frustration in magnetic systems occurs by competing exchange
interactions and leads frequently to disordered spin-liquid states
\cite{Nor09,Bal10,Lucile}. Recent progress in understanding transition
metal oxides with orbital degrees of freedom demonstrated many unusual
properties of systems with active $t_{2g}$ degrees of freedom --- they
are characterized by anisotropic hopping
\cite{Kha00,Har03,Dag08,Nic11,Wro10} which generates Ising-like orbital
interactions \cite{Jac04,Kha05,Jac07,Jac08,Kru09,Che09,Ryn10,Tro12,Che13},
similar to the
orbital superexchange in $e_g$ systems \cite{Dag04,Rei05}.
Particularly challenging are $4d$ and $5d$ transition metal oxides,
where the interplay between strong electron correlations and spin-orbit
interaction leads to several novel phases \cite{Wit14,Brz15}. In
iridates the spin-orbit interaction is so strong that spins and orbital
operators combine to new $S=1/2$ pseudospins at each site \cite{Jac09},
and interactions between these pseudospins decide about the magnetic
order in the ground state.

The $A_2$IrO$_3$ ($A$=Na, Li) family of honeycomb iridates has
attracted a lot of attention as these compounds have $t_{2g}$ orbital
degree of freedom and lie close to the exactly solvable $S=1/2$ Kitaev
model \cite{Kit06}. This model has a number of remarkable features,
including the absence of any symmetry breaking in its quantum Kitaev
spin-liquid (KSL) ground state, with gapless Majorana fermions
\cite{Kit06} and extremely short-ranged spin correlations \cite{Bas07}.
We emphasize that below we call a KSL also disordered spin-liquid
states which arise near the Kitaev points in presence of perturbing
Heisenberg interactions $\propto J$.

By analyzing possible couplings between the Kramers doublets it was
proposed that the microscopic model adequate to describe the honeycomb
iridates includes Kitaev interactions accompanied by Heisenberg exchange
in form of the Kitaev-Heisenberg (KH) model \cite{Jac09,Cha10}. Soon
after the experimental evidence was presented that several features of
the observed zigzag order are indeed captured by the KH model
\cite{Sin10,Liu11,Sin12,Cho12,Ye12,Comin,Gre13,Tro13,Cha13}. Its
parameters for $A_2$IrO$_3$ compounds
are still under debate at present \cite{Kat14,Val16}. One
finds also a rather unique crossover from the quasiparticle states to
a non-Fermi liquid behavior by varying the frustrated interactions
\cite{Tro14}.
Unfortunately, however, it was recently realized that this model does
not explain the observed direction of magnetic moments in Na$_2$IrO$_3$
and its extension is indeed necessary to describe the magnetic order
in real materials \cite{Chu15,Cha15}.
For example, bond-anisotropic interactions associated with the trigonal
distortions have to play a role to explain the differences between
Na$_2$IrO$_3$ and Li$_2$IrO$_3$ \cite{Rau15}, the two compounds with
quite different behavior reminiscent of the unsolved problem of
NaNiO$_2$ and LiNiO$_2$ in spin-orbital physics \cite{Rei05}.
On the other hand, the KH model
might be applicable in another honeycomb magnet $\alpha$-RuCl$_3$, see
e.g. a recent study of its spin excitation spectrum \cite{Ban16}.

Understanding the consequences of frustrated Heisenberg interactions on
the honeycomb lattice is very challenging and has stimulated several studies
\cite{Alb11,Cab11,Son16}. The KH model itself is highly nontrivial and
poses an even more interesting problem in the theory
\cite{Cha10,Cha13,Rau14,Oit15}: Kitaev term alone has
intrinsic frustration due to directional Ising-like interactions between
the spin components selected by the bond direction \cite{Kit06}.
In addition, these interactions are disturbed by nearest neighbor
Heisenberg exchange which triggers long-range order (LRO) sufficiently
far from the Kitaev points \cite{Cha10,Cha13,Rau14,Oit15}.
In general, ferromagnetic (FM) and antiferromagnetic (AF) interactions
coexist and the
phase diagram of the KH model is quite rich as shown in several previous studies \cite{Cha10,Cha13,Rau14,Oit15,Tre11,Ire14}.
Finally, the KH
model has also a very interesting phase diagram on the triangular
lattice \cite{Li15,Bec15,Jac15,Rou16}. These studies motivate better
understanding of quantum effects in the KH model on the honeycomb
lattice in the full range of its competing interactions.

The first purpose of this paper is to revisit the phase diagram of the KH
model and to investigate it further by combining exact diagonalization
(ED) result \cite{Cha13} with the self-consistent cluster mean field
theory (CMFT),
supplemented by the insights from the linear spin-wave (LSW) theory
and the second--order perturbation theory (SOPT).
The main advantage of CMFT is that it goes beyond
a single site mean field classical theory and gives not only the
symmetry-broken states with LRO, but also includes partly quantum
fluctuations, namely the ones within the considered clusters
\cite{Alb11,Brz12}. In this way the treatment is more balanced and may
allow for disordered states in cases when frustration of interactions
dominates. We present below a complete CMFT treatment of the phase
diagram which includes also the Kitaev term in MF part of the
Hamiltonian and covers the entire parameter space (in contrast to the
earlier prototype version of CMFT calculation on a single hexagon for
the KH model \cite{Got15}). Note that the CMFT complements the ED which
is unable to get symmetry breaking for a finite system, but nevertheless
can be employed to investigate the phase transitions in the present KH
model by evaluating the second derivative of the ground state energy to
identify phase transitions by its characteristic maxima \cite{Cha10,Cha13}.
ED result can be also used to recognize the type of magnetic order by
transforming to reciprocal space and computing spin-structure factor.
The second purpose is to investigate further the difference between
quantum KSL regions around both Kitaev points mentioned
in Ref. \cite{Cha13} and LRO/KSL boundaries.

The paper is organized as follows: In Sec. \ref{sec:KH} we introduce
the KH model and define its parameters. In Sec. \ref{sec:methods} we
present three methods of choice:
(i) the exact diagonalization in Sec. \ref{sec:ED},
(ii) the self-consistent CMFT in Sec. \ref{sec:CMFT}, and
(iii) linear spin wave theory in Sec. \ref{sec:lsw}.
An efficient method of solving the self-consistence problem obtained
within the CMFT is introduced in Sec. \ref{sec:lin}.
The numerical
results are presented and discussed in Sec. \ref{sec:res}:
(i) the phase transitions and the phase diagram are introduced in Sec.
\ref{sec:phd}, and
(ii) the phase boundaries, the values of the ground state energies
and the magnetic moments obtained by different methods are
presented and discussed in Secs. \ref{sec:qcen} and \ref{sec:qcom}, and
(iii) we discuss the compatibility of the Kitaev interaction with
different spin ordered states in Sec. \ref{sec:qcom}.
Spin correlations obtained for various phases are presented in
Sec. \ref{sec:ss}.
The dynamical spin susceptibility and spin structure factor are
presented for different phases in Sec. \ref{sec:chi}. Finally,
in Sec. \ref{sec:summa} we present the main conclusions and short
summary. The paper is supplemented with Appendix where we explain
the advantages of the linearization procedure implemented on the CMFT
on the example of a single hexagon.

\section{Kitaev-Heisenberg Model}
\label{sec:KH}

We start from the KH Hamiltonian with nearest neighbor interactions on
the honeycomb lattice in a form,
\begin{eqnarray}
{\cal H}&\equiv&
K\sum_{\langle ij\rangle\parallel \gamma} S_{i}^{\gamma} S_{j}^{\gamma}
+ J\sum_{\langle ij\rangle} \vc{S}_{i}\cdot \vc{S}_{j},
\label{ham_in}
\end{eqnarray}
where $\gamma=x$,$y$,$z$ labels the bond direction. The Kitaev term
$\propto K$ favors local bond correlations of the spin component
interacting on the particular bond. The superexchange $J$
is of Heisenberg form and alone would generate a LRO state,
antiferromagnetic or ferromagnetic, depending on whether $J>0$ or $J<0$.
We fix the overall energy scale, $J^{2}+K^{2}=1$,  and choose angular
parametrization
\begin{eqnarray}
\label{k}
K&=&
\sin\varphi,\\
\label{j}
J&=&
\cos\varphi,
\end{eqnarray}
varying $\varphi$ within the interval $\varphi\in[0,2\pi]$. This
parametrization exhausts all the possibilities for nearest neighbor
interactions in the KH model.

While zigzag AF order was observed in Na$_2$IrO$_3$
\cite{Sin12,Cho12,Ye12,Comin,Gre13}, its microscopic explanation has been
under debate for a long time.
The \textit{ab initio} studies \cite{Foy13,Kat15} give motivation to
investigate a broad regime of parameters $K$ (\ref{k}) and $J$ (\ref{j}).
Further motivation comes from the honeycomb magnet $\alpha$-RuCl$_3$
\cite{Ban16}. Note that we do not intend to identify the parameter sets
representative for each individual experimental system, but shall
concentrate instead on the phase diagram of the model Eq. (\ref{ham_in})
with nearest neighbor interactions only.

\section{Calculation methods}
\label{sec:methods}

\subsection{Exact diagonalization}
\label{sec:ED}

We perform Lanczos diagonalization for $N=24$-site cluster with
periodic boundary conditions (PBC). This  cluster respects all the
symmetries of the model, including hidden ones. Among the accessible
clusters it is expected to have the minimal finite-size effects.

\subsection{Cluster mean field theory}
\label{sec:CMFT}

A method which combines ED with  an explicit breaking of Hamiltonian's
symmetries is the so-called self-consistent CMFT. It has
been applied to several models with frustrated interactions, including
Kugel-Khomskii model \cite{Brz12}. The method was also extensively used
by Albuquerque \textit{et al.} \cite{Alb11} as one of the means to
establish the full phase diagram of Heisenberg-$J_{2}$-$J_{3}$ model on
the honeycomb lattice.

\begin{figure}[t!]
\includegraphics[width=7cm]{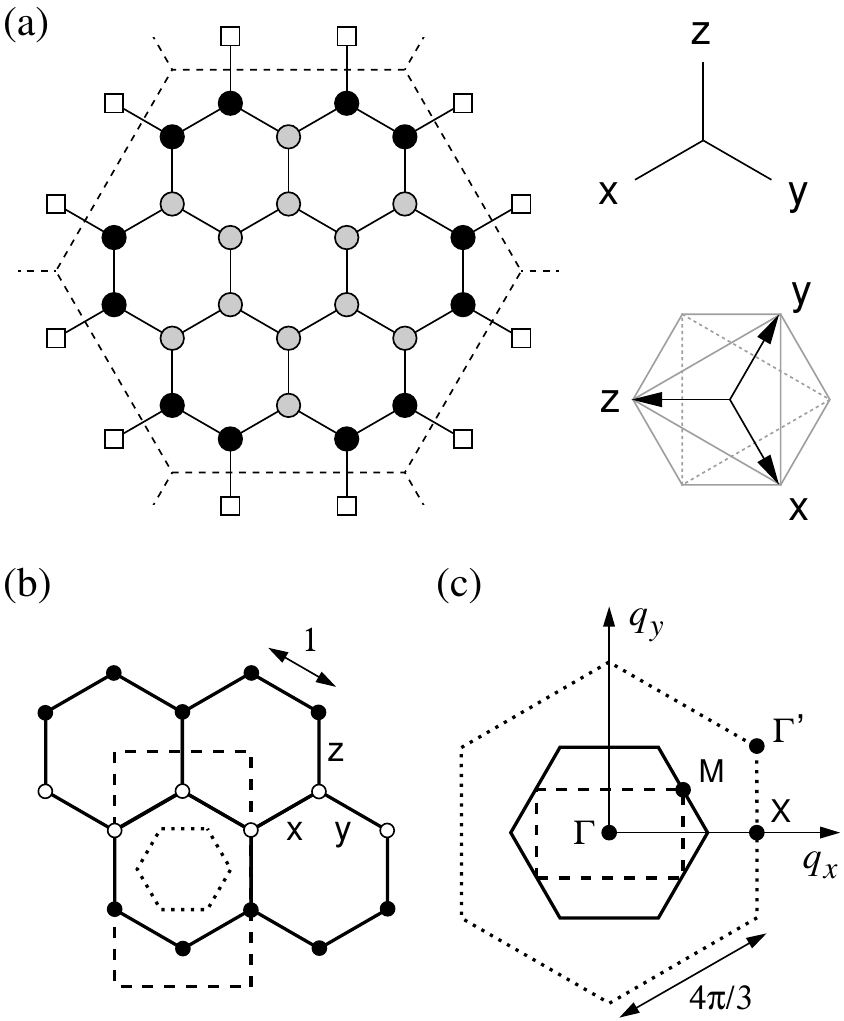}
\caption{(a) 24-site cluster and the introduction of the mean fields.
Gray (black) circles indicate internal (boundary) sites.
In CMFT  the internal bonds of the cluster are
treated exactly while the external bonds crossing the cluster boundary
(dashed) are treated on the MF level.
The sites marked by $\Box$ generate an effective magnetic fields
on the boundary sites $\bullet$.
Labels $x$, $y$ and $z$ stand for three inequivalent bond directions
determining the active products $S^{\gamma}_{i}S^{\gamma}_{j}$ in
Kitaev part of the Hamiltonian (\ref{ham_in}), e.g. bonds of $x$
direction contribute with the $S^x_iS^x_j$ product to the Hamiltonian.
The pseudospin axes used here are parallel to the cubic axes indicated
in the top view of a single octahedron.
(b) Unit cells:  for honeycomb lattice (coinciding with single hexagon
of that lattice), for triangular lattice (inner dotted hexagon) and
zigzag magnetic unit cell (dashed rectangle). Black and white circles
indicate one of three equivalent zigzag patterns.
(c) Corresponding Brillouin zones and special $\vc{q}$ points for the
lattice constant $a=1$.
}
\label{1}
\end{figure}

Within CMFT the internal bonds of the cluster [connecting the circles
in Fig. \ref{1}(a)] are treated exactly. The corresponding part
$H_{\rm IN}$  of the Hamiltonian is the nearest neighbor KH Hamiltonian,
Eq. \eqref{ham_in}.
The external bonds that connect the boundary sites ($\bullet$) with
the corresponding boundary sites of periodic copies of the cluster
($\Box$) are described by the MF part of the Hamiltonian,
\begin{eqnarray}
H_{\rm MF}&\equiv&
K\sum_{[ij]\parallel z} \langle S_{i}^{z}\rangle S_{j}^{z}
+ J\sum_{[ ij]} \langle S_{i}^{z}\rangle S^{z}_{j},
\label{ham_mf}
\end{eqnarray}
where $[ij]$ marks the external bonds. Since the ordered moments in KH
model align always along one of the cubic axes $x$, $y$, $z$ (see e.g.
Ref.~\cite{Cha10}) we have put
\begin{equation}
\langle\vec{S}_i\rangle\cdot\vec{S}_j\equiv\langle S^z_i\rangle S^z_j
\label{SzSz}
\end{equation}
in $H_{\rm MF}$ to simplify the calculations.

The averages $\langle S_i^z\rangle$ generate effective magnetic fields
acting on the boundary sites of the cluster. The total Hamiltonian
\begin{eqnarray}
{\cal H}&\equiv&
H_{\rm IN}+H_{\rm MF},
\label{ham_num}
\end{eqnarray}
is diagonalized in a self-consistent manner, taking slightly different
approach than the one presented in Ref. \cite{Alb11}:
instead of starting with random wave function our algorithm begins
with expectation values $\langle S_i^z\rangle_\mathrm{ini}$
on each boundary site $i$ of the cluster. These can represent a certain
pattern (zigzag, stripy, N\'eel, FM) or be  set randomly to have a
``neutral" starting point.
After diagonalizing the Hamiltonian
(\ref{ham_num}) (again by the ED Lanczos method) the ground state of
the system is obtained and we recalculate the expectation values
$\langle S^{z}_i\rangle$
to be used in the second iteration.
The procedure is repeated until self-consistency is reached.

\subsection{Linearized cluster mean field theory}
\label{sec:lin}

A single iteration of the self-consistent MF calculation may be viewed
as a nonlinear mapping of the set of initial averages
$\{\langle S_i^z\rangle_\mathrm{in}\}$ to the resulting averages
$\{\langle S_i^z\rangle_\mathrm{fin}\}$. The self-consistent solution is
then a stable stationary point of such a mapping. To find the leading
instability, we may consider the case of small initial averages in the
CMFT calculation and identify the pattern characterized by the fastest
growth during the iterations. To this end we linearize the above
mapping.

\begin{figure}[t!]
\includegraphics[width=8.2cm]{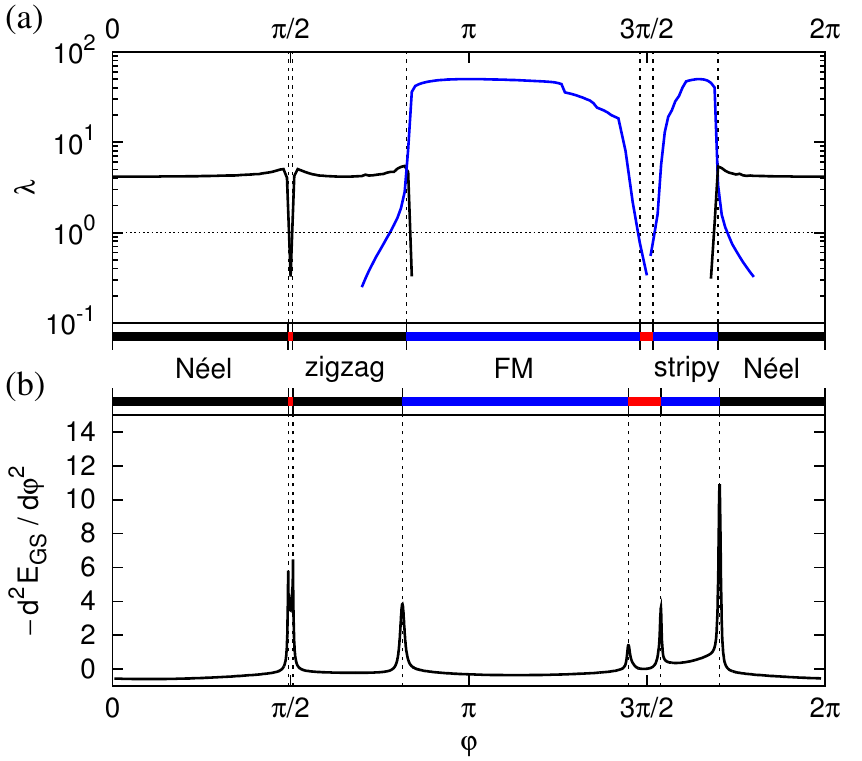}
\caption{
(a) The values of $\lambda$ obtained by the linearization
of CMFT for an embedded cluster of $N=24$ sites with fixed magnetic
order patterns: FM, AF, stripy, and zigzag.
Leading $\lambda >1$ indicates the order that sets in.
The disordered KSL phase is indicated by the red color.
(b) Second derivative of the ground state energy,
$-\mathrm{d}^2 E_0(\varphi)/\mathrm{d}\varphi^2$,
obtained by ED. Adopted from Ref.~\cite{Cha13}.
}
\label{la24}
\end{figure}

In the lowest order the mapping corresponds to the multiplication
of the vector of the averages $\{\langle S_i^z\rangle_\mathrm{in}\}$
by the matrix,
\begin{equation}
F_{ij} = \frac{\partial \langle S_i^z\rangle_\mathrm{fin}}
{\partial \langle S_j^z\rangle_\mathrm{in}},
\end{equation}
where $i$ and $j$ run through the cluster boundary sites. During iterations,
the patterns corresponding to the individual eigenvectors of the matrix
$F$ grow as $\lambda^n$, where $\lambda$ is a particular eigenvalue and
$n$ is the number of iterations. The ordering pattern obtained by CMFT
is then given by the eigenvector with largest $\lambda_{max}>1$. In the
quantum KSL regimes, all the eigenvalues are less than $1$
and no magnetic order emerges. An example of linearized CMFT applied to
a single hexagon with PBC can be found in the Appendix.

A modified version of this method, used to obtain Fig. \ref{la24}(a),
assumes a particular ordered pattern (N\'eel, zigzag, FM, or stripy
phase) and uses a single spin average $\langle S^z\rangle_\mathrm{in}$
distributed along the boundary sites outside the cluster, with the
signs consistent with this pattern. The resulting values,
$\langle S^z_i\rangle_\mathrm{fin}$, are then averaged correspondingly.
In this case the matrix $F$
is reduced to a single value $\lambda$ plotted in Fig. \ref{la24}(a).
We observe that the largest eigenvalue either drops below 1 when the
disordered KSL state takes over, or interchanges with
another eigenvalue corresponding to a different ordered phase.

\subsection{Linear spin-wave theory}
\label{sec:lsw}

The LSW method is a basic tool to determine
spin excitations and quantum corrections in systems with long-range
order \cite{Wal63}. For systems with coexisting AF and FM bonds quantum
corrections are smaller than for the N\'eel phase but are still
substantial for $S=1/2$ spins \cite{Rac02}. For the KH model the LSW
theory \cite{Cha10,Cha13,Cho12} has to be implemented separately for
each of the four ordered ground states: N\'eel (N),
zigzag (ZZ), FM, or stripy (ST). Then for a particular ground state
the Hamiltonian is rewritten in terms of the Holstein-Primakoff
bosons~\cite{Cho12,Mak15} and only quadratic terms
in bosonic operators are kept. The spectrum of such quadratic
Hamiltonian is finally obtained using the successive Fourier and
Bogoliubov transformations.

While the spin wave dispersion relations are usually of prime
interest~\cite{Cho12,Mak15,Cha10,Cha13}, there are also two other quantities
which can easily be calculated using LSW and which will be important
in the discussion that follows:
(i) the value of the total ordered moment $\langle M\rangle$ per site, and
(ii) the total energy per site $\langle E\rangle$.
These observables are calculated in a standard way~\cite{Wal63,Rac02}
and expressed in terms of the eigenvalues, i.e., spin-wave
energies $\omega_{{\bf k}\alpha}$, and the eigenvector components ($v_{{\bf k}
\alpha \lambda}$) of the bosonic
Hamiltonian {\it before} the Bogoliubov transformation:
\begin{align}
\langle M \rangle =S- \frac{1}{L V} \sum_{\alpha, \lambda = 1, ..., L}
\int_{{\bf k} \in BZ} |v_{{\bf k} \alpha, \lambda}|^2\ d^2{\bf k},
\label{m}
\end{align}
and
\begin{align}
\langle E \rangle =&  E_{\rm class}\  [S^2 \rightarrow S(S+1)] \nonumber \\
&+ \frac{S}{2 L V} \sum_{\alpha = 1, ..., L}  \int_{{\bf k} \in BZ}
{\omega_{{\bf k} \alpha}}\ d^2{\bf k},
\label{e}
\end{align}
where the choice of the sign of the eigenvalues and the normalization
of their eigenvectors is described in Ref.~\cite{Wal63}. Here
$E_\mathrm{class}$ is the classical ground state energy per site, e.g.
\begin{equation}
E_{\rm class} = -JzS^2/2,
\end{equation}
with $z=3$ for the N\'eel phase at $K=0$ and $S=1/2$ is the value of
spin quantum number. $L$ in Eqs. (\ref{m})-(\ref{e}) is the number of
the eigenvalues of the problem (spin-wave modes) and $\alpha$ enumerates
these modes. For all cases except for the zigzag order~\cite{Cha10},
the integrals go over the two-sublattice ($L=2$) rectangular Brillouin
zone (BZ)~\cite{Wei91} with its volume $V=8\pi^2 / 3\sqrt{3}$ and
$ -\pi / \sqrt{3} \le k_x \le \pi / \sqrt{3}$,
$-2 \pi /3 \le k_y \le 2 \pi / 3 $
(as already mentioned we assume the lattice constant $a=1$).
For the zigzag state $L=4$ and the rectangular BZ can be chosen as:
$ -\pi/\sqrt{3} \le k_x \le \pi/\sqrt{3}$ and
$- \pi / 3 \le k_y \le  \pi / 3$
and its volume is $V=4 \pi^2 / 3\sqrt{3}$.

\section{Quantum phase transitions}
\label{sec:res}

\subsection{Phase diagram}
\label{sec:phd}

\begin{figure}[t!]
\includegraphics[width=8.2cm]{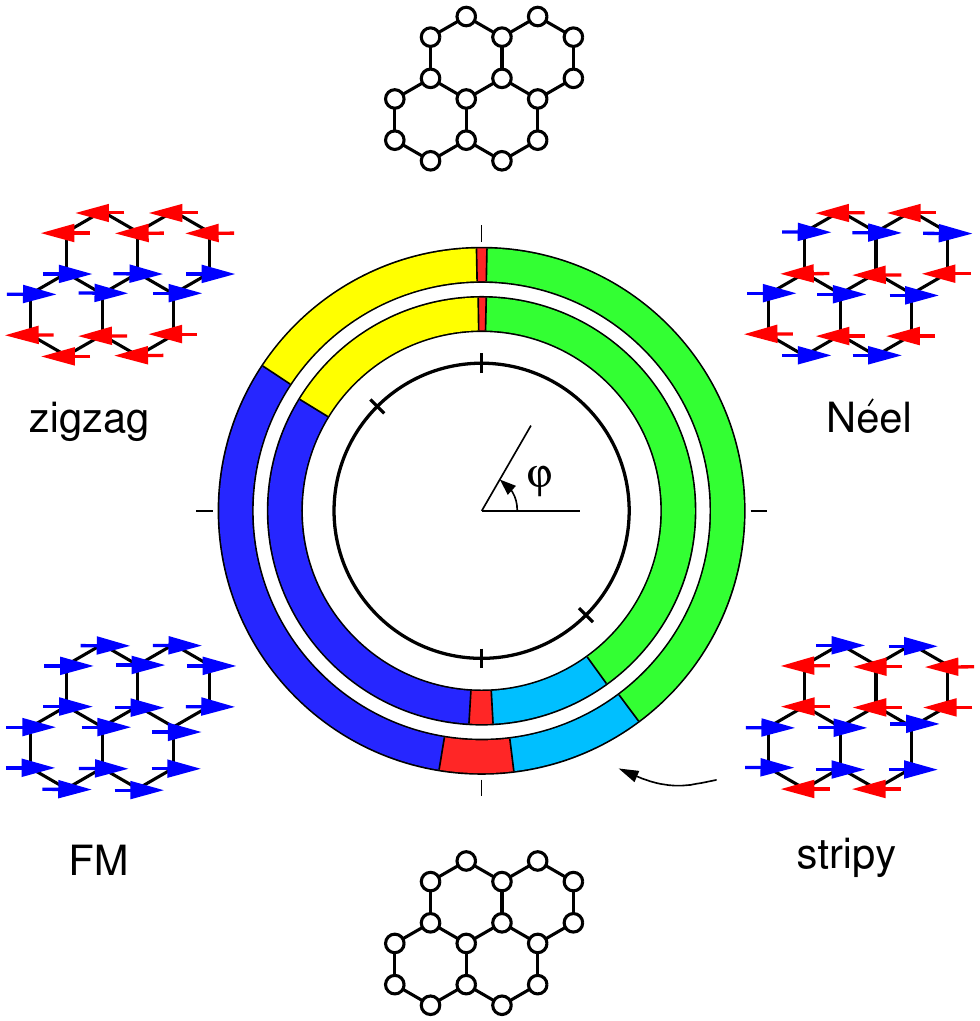}
\caption{
T$=0$ phase diagram for KH model. The outer ring is composed
from ED data for the 24-site cluster, reproducing the result from
Ref. \cite{Cha13} in the new parametrization, the middle ring shows
CMFT results also for 24-site cluster and the inner black circle
represents the classical result. The convention used
for the angular parameter $\varphi$ which determines
coupling constants [see Eqs. (\ref{k}) and (\ref{j})] is shown in the
center of the inner circle.  The colors represent particular phases,
shown also as mini-drawings next to suitable regions of the phase
diagram. Starting from $\varphi=0$ green colored region corresponds to
N\'eel order, red --- KSL, yellow --- zigzag order,
dark blue --- FM, red --- KSL, light blue --- stripy
phase and again green --- N\'eel phase.
}
\label{phase_diag}
\end{figure}

Here we supplement the ED--based phase diagram for the KH model established
in Ref.~\cite{Cha13} with the one obtained within CMFT.
Figure \ref{phase_diag} displays the phase boundaries obtained with ED
\cite{Cha13}, within CMFT, as well as classical (Luttinger-Tisza) phase
boundaries. The latter are included for completeness and
to highlight the fact that the quantum fluctuations stabilize the KSL
phases beyond single points, see below. To examine them in
more detail it is instructive to analyze the data in
Fig.~\ref{la24}(a) for the boundaries obtained from linearized CMFT and
Fig.~\ref{la24}(b) for the peaks in the second derivative of energy,
$-\mathrm{d}^2 E_0(\varphi)/\mathrm{d}\varphi^2$,
giving phase boundaries in ED~\cite{Cha13}.

It is clearly visible that all the methods that include quantum
fluctuations give quantum versions of the four classically established
magnetic phases: N\'eel, zigzag, FM and stripy. As the most important
effect we note that when quantum fluctuations are included within a
classical phase, the energy is generally lowered and that the emerging
phase is expected to expand beyond the classical boundaries, but only
in case if a phase which competes with it has weaker quantum
fluctuations. This implies that phases of AF nature will expand at the expense
of the FM ones as the latter phases have lower energy gains by quantum
fluctuations (which even vanish exactly for the FM order at $K=0$ and
$J<0$).

We summarize the phase boundaries obtained within different methods
in Table \ref{table1}. One finds substantial corrections to the quantum
phase transitions which follow from quantum fluctuations. These
corrections are quite substantial in both KSLs at the Kitaev
points ($K=+1$, $\varphi=\frac{1}{2}\pi$ and $K=-1$,
$\varphi=\frac{3}{2}\pi$, first column of Table~\ref{table1}). Indeed,
in the classical approach massively degenerate ground states exist just
at isolated points but they are replaced by disordered spin-liquid states
that extend to finite intervals of $\varphi$ when
quantum fluctuations are included, see the second, third and fourth
column in Table \ref{table1}. The expansion of N\'eel and zigzag phases
beyond classical boundaries is given by particularly large corrections
and is well visible.	

The most prominent feature in the phase diagram described above is
however the difference in size between two KSL regions, already
addressed before using ED \cite{Cha13} and also visible now in the CMFT
data. Therefore, the CMFT result supports the claim from Ref.
\cite{Cha13} that the stability of KSL perturbed by
relatively small Heisenberg interaction depends on the nature of the
phases surrounding the spin liquid and the amount of quantum
fluctuations that they carry.
In the following we discuss the above issues more thoroughly, examining:
(i) ground state energy curves emerging from ED, CMFT, SOPT within the linked
cluster expansion and LSW, (ii) the ordered moment given by various methods,
(iii) the spin--spin correlation functions, and (iv) the spin structure factor
as well as the dynamical spin susceptibility in the vicinity of the Kitaev
points.

\subsection{Quantum corrections: energetics}
\label{sec:qcen}

\begin{figure}[t!]
\includegraphics[width=8.4cm]{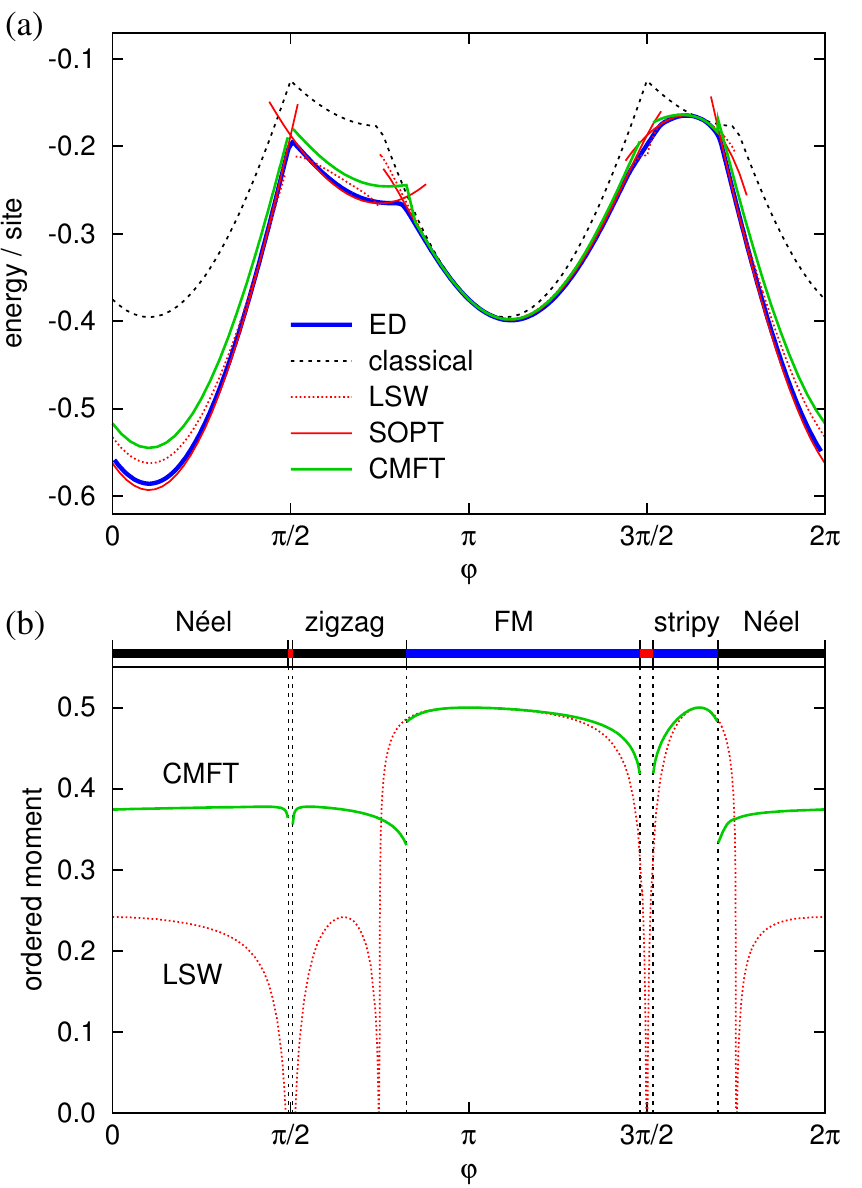}
\caption{
(a) Comparison between ground state energies per site obtained
using various methods:
classical Luttinger-Tisza approximation (dashed black),
SOPT (solid red),
LSW approximation (dashed red),
ED for 24-site cluster (solid blue, see \cite{Cha13} for this result
in a different parametrization),
and CMFT (solid green).
(b) Ordered moment value obtained from CMFT (green line) and LSW
(dashed red line). For CMFT the values were obtained by averaging
$\pm\langle S_i^z\rangle$ over the boundary sites, with the signs
determined by the particular type of magnetic order.
}
\label{sz24}
\end{figure}

We start the discussion of quantum corrections to the energy of the ordered
phases by noting that, even though it properly captures finite order
parameters, the CMFT looses quantum energy on the external bonds and does not
therefore provide a reliable estimate of the ground-state energy. Instead, the
energy obtained using the ED calculations [see Fig. \ref{sz24}(a)] will be
treated as a reference value. This is supported by the fact that the ED phase
boundaries were roughly confirmed by tensor networks (iPEPS) \cite{Ire14} and
DMRG results \cite{Tre11}:
While the iPEPS phase
boundaries agree with ED for AF KSL/LRO transitions and
the boundaries between different LRO phases differ only slightly from
those found in ED (iPEPS: zigzag/FM -- $0.808\pi$, stripy/N\'eel --
$1.708\pi$), for FM KSL/LRO transition however the iPEPS result
KSL/stripy -- $1.528\pi$).
On the other hand, DMRG boundaries agree perfectly with ED and due to
four--sublattice dual transformation~\cite{Kha05,Cha10} one can
reproduce the FM/zigzag as well as FM/KSL boundaries. Only the
extent of the AF spin-liquid phase cannot be extracted from this result,
but that is already confirmed by iPEPS.

\begin{table}[t!]
\caption{Phase boundaries for KH model, parameterized by the angle
$\varphi$ (in units of $\pi$), see Eqs. (\ref{k}) and (\ref{j}).
Columns: classical Luttinger-Tisza approximation,
second--order perturbation theory,
exact diagonalization, and
self--consistent cluster mean field theory. }
\begin{ruledtabular}
\begin{tabular}{ccccc}
 boundary & classical & SOPT & ED & CMFT \cr
\hline
N\'eel/KSL  &$0.5$&$0.492$&$0.494 $&$0.493 $  \cr
KSL/zigzag&$0.5$&$0.507$&$0.506 $&$0.505$\cr
zigzag/FM&$ 0.75$&$0.813$&$0.814$ &$0.825$\cr
FM/KSL&$1.5$ &$1.463$&$1.448$&$1.481$\cr
KSL/stripy& $1.5$&$1.530$&$1.539$&$1.517$\cr
stripy/N\'eel&$1.75$&$1.705$&$1.704$&$1.699$\cr
\end{tabular}
\end{ruledtabular}
\label{table1}
\end{table}

Fig. \ref{sz24}(a) shows a quite remarkable agreement between the energy
values and critical values of $\varphi$ obtained by the simplest SOPT
\cite{Cha10} and our reference ED results.
This suggests that this analytical method can be utilized to get better
insight to the quantum contributions to the ground state energy.
For the four phases with LRO, the energy per site ${\cal E}$, written as a sum of
the classical energy $E_\mathrm{class}$ and the quantum fluctuation
contribution $\Delta E$, is obtained as:
\begin{eqnarray}
\label{enen}
{\cal E}_{\rm N} &=&-\frac18 (K+3J) -\frac{1}{16}(K+3J),   \\
\label{enez}
{\cal E}_{\rm ZZ}&=&-\frac18 (K-J)  -\frac{1}{16}(K-J) ,   \\
\label{enef}
{\cal E}_{\rm FM}&=&+\frac18 (K+3J) +\frac{1}{16} \frac{K^2}{K+2J}, \\
\label{enes}
{\cal E}_{\rm ST}&=&+\frac18 (K-J)  +\frac{1}{16}\frac{(K+2J)^2}{K}.
\end{eqnarray}
In addition, to get the LRO/KSL phase boundary points in Table \ref{table1},
we estimate the energy of the KSL phase as
\begin{equation}
{\cal E}_{\rm KSL}\simeq
\frac{3}{2}(K+J)\langle S^\gamma S^\gamma\rangle_\mathrm{Kitaev},
\end{equation}
using the analytical result for the Kitaev points \cite{Bas07},
$\langle S^\gamma S^\gamma\rangle_\mathrm{Kitaev}\approx\pm 0.131$.

The two spin-liquid phases in the phase diagram of KH model differ
strongly in their extent, despite the formal equivalence of the FM
and AF Kitaev points provided by an exact mapping of the Hamiltonian
\cite{Kit06}. As mentioned earlier, this is due to the fact that the
two KSLs compete with LRO phases of a distinct nature. Here we
give a simple interpretation based on the strength of the quantum corrections
of the LRO phases estimated using \eqref{enen}--\eqref{enes}.
Later, in Secs. \ref{sec:ss} and \ref{sec:chi} we illustrate the different
nature of the transitions between FM and AF KSL and the surrounding it
LRO phases in terms of spin correlations and spin dynamics.

Let us now compare the quantum fluctuation contribution and the
classical one. For the LRO phases surrounding the AF spin liquid ---
N\'eel and zigzag --- we always have
$\Delta E/E_\mathrm{class}=\frac{1}{2}$ as deduced from Eqs.
(\ref{enen}) and (\ref{enez}), i.e., only $\frac23 {\cal E}_{\rm N}$
and $\frac23 {\cal E}_{\rm ZZ}$ are found in the classical approach.
This guarantees that the quantum phase transition between these two
types of order occurs at the same value of $\varphi=\pi/2$ in SOPT and
in the classical approach that do not capture the spin-liquid phase in
between these ordered states, see Fig.~\ref{sz24}(a).
In contrast, the phases neighboring to the FM
spin liquid --- FM and stripy --- would reach the value of
$\Delta E/E_\mathrm{class}=\frac12$ only at the FM Kitaev point with
$J=0$ and away from this point the contribution of quantum fluctuations
decreases rapidly allowing for large extent of the FM spin-liquid phase.
Note, that both these latter phases contain a point which is exactly
fluctuation free --- for FM phase when frustration is absent ($K=0$),
and for stripy phase it is related to the FM one by the interaction
transformation \cite{Cha15} at $K=-2J$.

Moving to the CMFT energy analysis (green line in Fig.~\ref{sz24}(a))
one should also keep in mind that within the CMFT method the external
bonds between $\langle S^z_i\rangle$ and $S^z_j$ do not include quantum
fluctuations fully. This implies worse estimate of the energy for
regions of the phase space that allow quantum fluctuations.
As a consequence the region of stability of FM spin-liquid phase is
smaller than that obtained in the ED. Finally, the estimates obtained
from LSW, which represents a harmonic approximation to the quantum
fluctuations, are typically better than those from the CMFT but not as
good as those from SOPT, see dashed red lines in Fig. ~\ref{sz24}(a).
As expected, the LSW energy fits well with ED curve for FM and stripy
phases with less quantum fluctuations and starts to diverge when
beyond quantum phase transitions within N\'eel and zigzag phases.

\subsection{Quantum corrections: ordered moment}
\label{sec:qcom}

As usual,
getting the correct value of the ordered moment turns out to
be a more difficult task than estimating the ground state energy. This
is primarily due to the fact that the ED does not capture the
symmetry-broken states and the ordered moment can only be indirectly
extracted from the $m^2$; moreover, the SOPT may not be reliable here.
Hence, we are mostly left with the results obtained with CMFT and LSW.
We discuss the corresponding data [shown in Fig.~\ref{sz24}(b)]
together with the several values given already in the literature.

Let us begin with the Heisenberg AF point  $\varphi=0$: here it is
expected that the ordered moment should be
strongly reduced by quantum fluctuations.
LSW approximation estimates the ordered moment value at $0.248$ \cite{Wei91}.
Similar values were extracted from $m^2$ in quantum Monte Carlo
($0.268$ \cite{Reg89,Cas06,Low09}) and ED calculations ($0.270$ \cite{Alb11}).
In the last case however the authors admit that the set of clusters
for finite size scaling was chosen so as to make the best agreement
with quantum Monte Carlo.
Another method --- series expansion (high order perturbation theory)
\cite{Oit15} sets ordered moment value at a somewhat higher value of
$0.307$. While all the above results seem roughly consistent, CMFT
value seems to stand out ($0.374$ for $\varphi=0$). Nevertheless,
one should note that the ordered moment estimated from $m^2$ for
24--site cluster ED equals $0.45$ \cite{Alb11} which is above the CMFT value.
This suggests that at this point the finite size scaling is important.

Before transferring to the frustrated regime we briefly mention that
the the trivial ordered moment value at $\varphi=\pi$ is here correctly
reproduced by both CMFT and LSW. Besides, for the regions around the
fluctuation--free FM (and stripy) point the
ordered moments predicted by CMFT and LSW also match. Following the
ground state energy analysis, LSW gives the correct result because
quantum fluctuations contribution is small compared to the classical
state. The further we move towards the Kitaev points, however, the more
incorrect the LSW approximation should be because of the strong
reduction of the ordered moment due to the growing frustration.

In contrast, the lack of quantum fluctuations on the external bonds
makes CMFT steadily biased except for FM and stripy phases.
However, since for the internal part of the cluster the fluctuations
are still fully included, the frustration should be well handled and
CMFT should give more predictable results than LSW in frustrated parts
of the phase diagram.
Here it is also important to stress, that the series expansion captures
correctly the fluctuation--free point at $\varphi=\pi$ (FM) and
$\varphi=-\arctan 2$ (stripy) and predicts a broader region of FM
KSL phase \cite{Oit15}. The order parameter is also
qualitatively correctly estimated and is reduced more to $m\simeq 0.3$
for both N\'eel and zigzag phases \cite{Oit15}. However, while the
ordered moment values obtained by CMFT are consistent with the
four--sublattice dual transformation, the ordered moment data from the
high--order perturbation theory \cite{Oit15} are not, as the ordered
moment values differ at the points connected by the mapping.
Unfortunately the largest difference appears near the FM LRO/KSL
boundaries. This observation uncovers certain limits of the
high--order perturbation theory.

\subsection{Quantum corrections: naive interpretation}
\label{sec:naive}

Let us conclude the discussion of the quantum corrections with the
following more general observation:
Developing the argumentation presented by Iregui, Corboz, and Troyer
\cite{Ire14}, the dependence of the quantum correction to the energy
and to the ordered moment on the angle $\varphi$ suggests that the
Kitaev interaction is less ``compatible'' with the FM/stripy ground
states than with the N\'eel/zigzag ones. This can be understood in
the simple picture of the KH model on a 4-site segment of the honeycomb
lattice consisting of three bonds attached to a selected lattice site,
as presented below.

Starting with $\varphi=\pi$  (FM ground state. e.g. along the $z$
quantization axis) and increasing $\varphi$ leads to an addition of the
FM Kitaev term, which favors FM-aligned spins along the $x$, $y$, and
$z$ quantization axes for the $x$, $y$, and $z$ directional bonds,
respectively. It can easily be seen that, e.g. for the $x$ bond, the
eigenstate of the FM Kitaev-only Hamiltonian on that bond
($|\uparrow_x \uparrow_x \rangle$) has a $25\%$ overlap with the FM
ground state,
$|\langle \uparrow_z\uparrow_z|\uparrow_x\uparrow_x\rangle|^2=\frac14$.
In contrast, while a similar situation happens for the $y$ bond,
for the $z$ bond there is a $100\%$ overlap between such states.

Next, we perform a similar analysis for $\varphi=0$ and firstly assume
that we have a {\it classical} ground state. In this case for the
``unsatisfied'' bonds from the point of view of the increasing AF
Kitaev interaction we also obtain that the eigenstate of the AF
Kitaev-only Hamiltonian ($|\uparrow_x\downarrow_x\rangle$) on that bond
has a $25\%$ overlap with the classical N\'eel ground state --- e.g.:
$|\langle\uparrow_z\downarrow_z|\uparrow_x\downarrow_x\rangle|^2=
\frac{1}{4}$.  However, this situation changes once we consider that
the spin quantum fluctuations dress the classical N\'eel ground state.
This can be best understood if we assumed the unrealistic but
insightful case of very strong quantum fluctuations destroying the
classical N\'eel ground state: then for the $x$ bond a singlet could be
stabilized and the overlap between such a state and the state
``favored'' by the Kitaev term increases to $50\%$:
$|\langle 0 | \uparrow_x \downarrow_x \rangle |^2=\frac{1}{2}$.
This suggests that the N\'eel ground state, which {\it contains}
quantum spin fluctuations, is more ``compatible'' with the states
``favored'' by the Kitaev terms than the FM ground state, resulting in
more stable values of ordered moment for N\'eel phase.
It seems that the above difference is visible in CMFT data but not in
LSW ones. We shall discuss this issue further by analyzing spin
correlations below.

\section{Spin correlations}
\label{sec:ss}

\begin{figure}[t!]
\includegraphics[width=8.2cm]{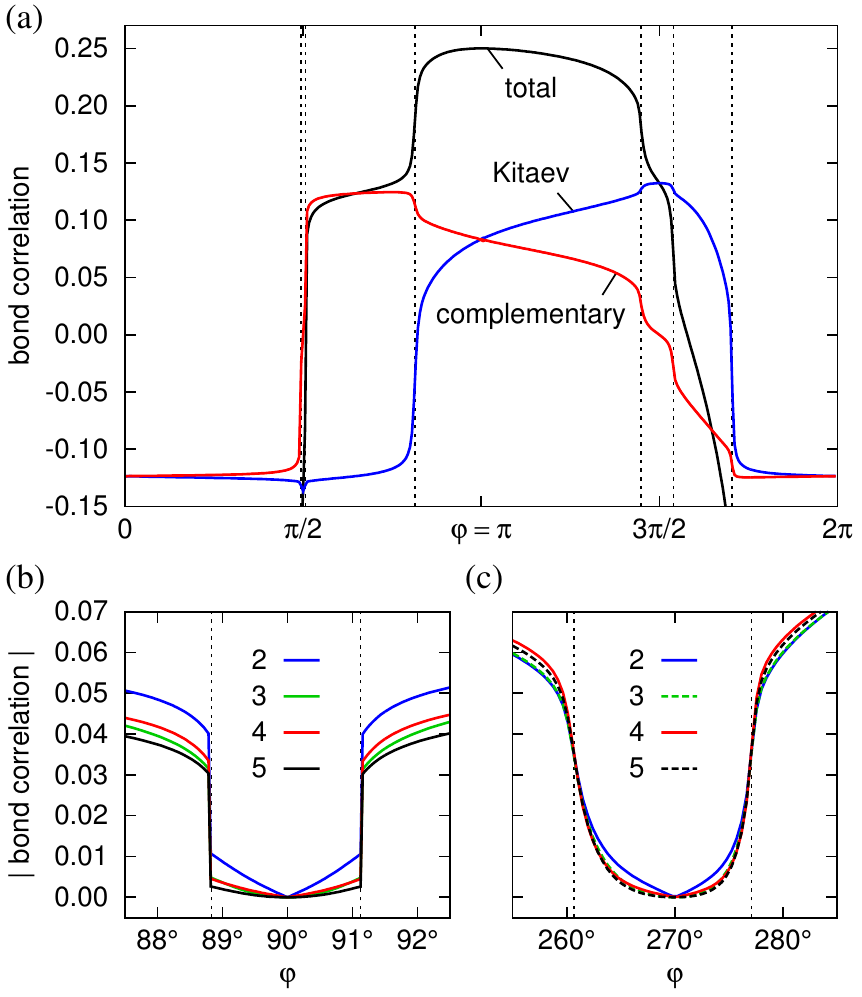}
\caption{
(a) Spin correlations $\langle\vc S_i\cdot\vc S_j\rangle$ obtained
within ED for the bonds between nearest neighbors (black line),
spin correlations of the components active in the Kitaev interaction,
$\langle S^\gamma_iS^\gamma_j\rangle$ (blue line), and complementary
spin components, $\langle S^{\bar\gamma}_i S^{\bar\gamma}_j\rangle$
(red line). Below further neighbor spin correlations
$|\langle \vc S_i\cdot \vc S_j\rangle|$ are shown:
(b) near the AF spin-liquid phase, and
(c) for the angle $\varphi$ interval including the FM spin-liquid phase.
}
\label{cor}
\end{figure}

Additional information about the ground state is given by spin--spin
correlation functions. In  Fig. ~\ref{cor}(a) one can observe
isotropic stable $\langle S^{\gamma}_{i}S^{\gamma}_{j}\rangle$
correlations in almost the entire AF phase
($\langle\boldsymbol{S}_{i}\cdot\boldsymbol{S}_{j}\rangle\approx-0.36$
for $\varphi=0$), while for FM phase the anisotropy quickly develops
when moving away from FM Heisenberg point $\varphi=\pi$ (here
$\langle\boldsymbol{S}_{i}\cdot\boldsymbol{S}_{j}\rangle$ reaches the
classical value $0.25$). This again demonstrates that the AF (and
zigzag) phase is more robust and uniform than FM (and stripy) phase.

Moreover, spin-spin correlations allow us to confirm the disordered
regions around the Kitaev points as critical cases of quantum spin
liquid \cite{Tik11}. At the Kitaev points we observe the expected
undisturbed KSL pattern: non--zero values of nearest
neighbor correlations between spin components active in the Kitaev
interaction (blue curve in Fig. ~\ref{cor}(a)) and vanishing
correlations between complementary components (red curve). In contrast,
the next nearest and further neighbor correlations disappear, see Figs.
\ref{cor}(b) and \ref{cor}(c). While moving away from the Kitaev points
the absolute values of the correlations enter the regions of slow
growth --- these are signatures of the critical spin-liquid phases and
they look similar in AF and FM spin liquid cases. At some point however
proceeding further results in rapidly growing absolute values which
mark KSL/LRO boundaries.

Furthermore, Figs. \ref{cor}(b) and \ref{cor}(c) prove that there is
a qualitative difference between the two spin-liquid regimes. This is
observed in the rapid growth of spin correlations at the onset of LRO:
step-like jump visible in Fig. \ref{cor}(b) contrasts with smoother
crossover seen in Fig. \ref{cor}(c). Below we investigate this distinct
behavior by analyzing the dynamical spin susceptibility for various
available phases. After Fourier transformation of the $z$--component
correlations, we obtain the spin structure factor to be discussed in
the context of the spin susceptibility also in Sec.~\ref{sec:chi}.

\section{Spin susceptibility and excitations
         in the vicinity of the Kitaev points}
\label{sec:chi}

Below we study the spin dynamics within the KH model by
analyzing the dynamical spin susceptibility at $T=0$,
\begin{equation}
\chi_{\alpha\alpha}(\vc q,\omega) = i \int_0^\infty \left\langle\Phi_0|
[S^\alpha_{\vc q}(t),S^\alpha_{-\vc q}(0)]
\right|\Phi_0\rangle\, \mathrm{e}^{i\omega t} \,\mathrm{d} t,
\end{equation}
with the Fourier-transformed spin operator defined via
\begin{equation}
S^\alpha_{\vc q} = \frac1{\sqrt{N}}
\sum_{\vc R} \mathrm{e}^{-i\vc q\cdot\vc R} S^\alpha_{\vc R}
\;
\end{equation}
and $|\Phi_0\rangle$ denoting the cluster ground state. For $\omega>0$,
the imaginary part of $\chi(\vc q,\omega)$ reads as
\begin{equation}
\chi''_{\alpha\alpha}(\vc q,\omega) = -\mathrm{Im}\,
\langle\mathrm{\Phi_0}|\, S^\alpha_{\vc q} \;
\frac1{\omega+E_\mathrm{GS}-\mathcal{H}+i\delta}
\; S^\alpha_{-\vc q} \,
|\mathrm{\Phi_0}\rangle \;,
\end{equation}
which can be conveniently expressed as a sum over the excited states
$\{|\nu\rangle\}$,
\begin{equation}\label{eq:chiexcsum}
\chi''_{\alpha\alpha}(\vc q,\omega) = \pi\sum_{|\nu\rangle}
 |\langle\nu|S^\alpha_{-\vc q}|\Phi_0\rangle|^2
\delta(\omega-E_\mathrm{\nu}) \;,
\end{equation}
where the excitation energy $E_\nu$ is measured relative to the ground
state energy $E_{\mathrm{GS}}$. We have evaluated
$\chi_{\alpha\alpha}(\vc q,\omega)$ by ED on a hexagonal cluster of
$N=24$ sites. In the ED approach, the exact ground state of the cluster
$|\Phi_0\rangle$ is found by Lanczos diagonalization, the operator
$S^\alpha_{-\vc q}$ is applied, and the average of the resolvent
$1/(z-\mathcal{H})$ is determined by Lanczos method using normalized
$S^\alpha_{-\vc q}|\Phi_0\rangle$ as a starting vector \cite{Ful95}.

\begin{figure}[t!]
\includegraphics[width=8.4cm]{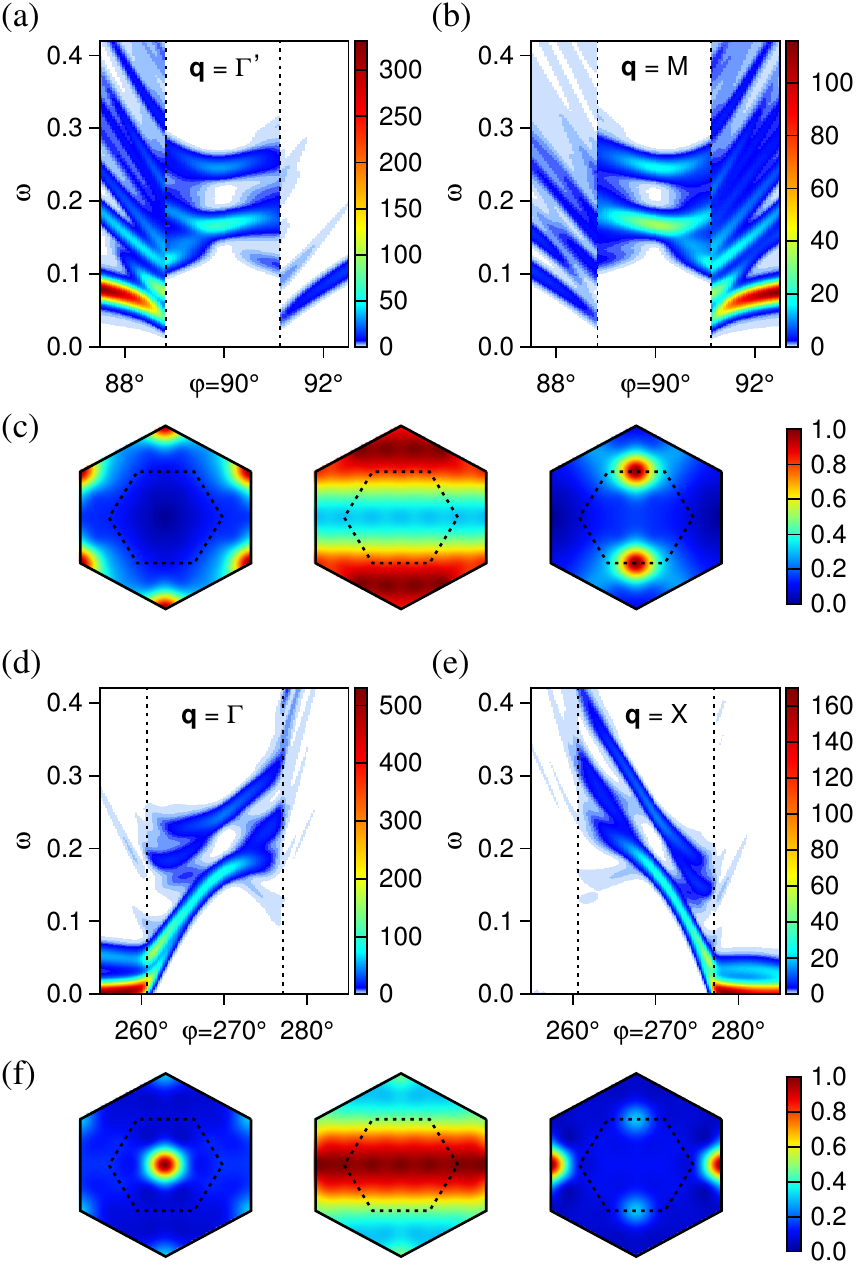}
\caption{
(a)~Dynamical spin susceptibility $\chi''(\vc q,\omega)$ obtained by
ED near the AF KSL phase at the characteristic wavevector of the AF
order, $\vc q=\Gamma'$.
(b)~The same for the zigzag wavevector $\vc q=M$.
(c)~Brillouin zone portraits of the spin-structure factor
$\langle S^z_{-\vc q} S^z_{\vc q}\rangle$ at $\varphi=87.5^\circ$,
$90^\circ$, and $92.5^\circ$ (interpolated from the ED data). The inner
hexagon is the Brillouin zone of the honeycomb lattice, the outer one
corresponds to the triangular lattice with the missing sites filled in.
(d,e)~The same as in panels (a,b) but for the interval containing the
FM ($\vc{q}=\Gamma$) and stripy ($\vc q=X$) phase.
(f)~Brillouin zone portraits of the spin-structure factor obtained at
$\varphi=255^\circ$, $270^\circ$, and $285^\circ$.
}
\label{susc}
\end{figure}

In our case of the KH model, the calculation generally requires a
relatively large number of Lanczos steps (up to one thousand) to
achieve convergence of the dense high-energy part of the spectrum.
Having the advantage of being exact, the method is limited by the
$\vc q$ vectors accessible for a finite cluster and compatible with
the PBC, and by finite-size effects due to small $N$. These concern
mainly the low-energy part of $\chi''$ and lead e.g. to an enlarged
gap of spin excitations in LRO phases of AF nature. Nevertheless,
a qualitative understanding can still be obtained.

The evolution of numerically obtained $\chi''$ with varying $\varphi$
is presented in Figs.~\ref{susc}(a) and \ref{susc}(b) for the region
including AF spin-liquid phase, as well as in Figs.~\ref{susc}(d) and
\ref{susc}(e) for the region including the FM spin-liquid phase.
The transitions are well visible at the characteristic $\vc q$ vectors
of the individual LRO phases. The structure factor pattern, see
Figs.~\ref{susc}(c) and \ref{susc}(f), changes accordingly between the
sharply peaked one in LRO phases and a wave-like form characteristic
for nearest neighbor correlations in the spin-liquid phases.

After entering the spin-liquid phase, further changes of the spin response are
very different for the AF and FM case. In the AF case, there is a sharp
transition --- a level crossing at our cluster, so that the ground state
changes abruptly. The original intense pseudo-Goldstone mode as well as many
other excited states become inactive in the spin-liquid phase. The observed
low--energy gap in $\chi''$ varies only slightly with $\varphi$.

In contrast, when entering the FM spin-liquid phase the excitation
that used to be the gapless magnon mode is characterized by a
gradually increasing gap which culminates at the Kitaev point.
Starting from the Kitaev point, the gradual reduction of the
low--energy gap in $\chi''$ due to the Heisenberg perturbation
manifests itself by a development of spin correlations beyond nearest
neighbors (already reported in Fig.~2 of Ref.~\cite{Cha10}) and an
increase of the static susceptibility to the magnetic field
Zeeman-coupled to the order parameter of the neighboring LRO phase.
This susceptibility then diverges at the transition point
(see also Fig.~3 of Ref.~\cite{Cha10}).

\section{Summary and conclusions}
\label{sec:summa}

In the present paper we studied the phase diagram of the Kitaev-Heisenberg
model by a combination of exact diagonalization and cluster mean field theory
(CMFT), supplemented by the insights from linear spin-wave theory and the
second--order perturbation theory.  Both methods allowed to stabilize
previously known ordered phases: N\'eel, zigzag, FM and stripy.  Moreover, the
ordered moment analysis provided by cluster mean field approach demonstrates
N\'eel--zigzag and FM--stripy connections described before \cite{Cha13}.
Compared to the previous CMFT studies utilizing $N=6$ site cluster (see Ref.
\cite{Got15} or the Appendix), we have used a sufficiently large cluster of
$N=24$ sites preserving the lattice symmetries and improving the ratio between
internal and boundary bonds. This led to a balanced approach which allowed us
to treat both ordered and disordered (spin-liquid) states on equal footing.

As the main result, the present study uncovers a fundamental difference
between the onset of broken symmetry phases in the vicinity of Kitaev
points with antiferromagnetic or ferromagnetic interactions. While the
spin liquids obtained at $K=+1$ and $K=-1$ are strictly equivalent and
can be transformed one into the other in the absence of Heisenberg
interactions (at $J=0$), spin excitations and quantum phase transitions
emerging at finite $J$ are very different in both cases.
For antiferromagnetic Kitaev spin liquid phase ($K\simeq1$) one finds
that a gap opens abruptly in $\chi''(\vc q,\omega)$ at
$\vc q=\Gamma^{'}$ and $\vc q=M$ when the ground state changes to the
critical Kitaev quantum spin liquid. This phase transition is abrupt
and occurs by level crossing. In contrast, for ferromagnetic spin liquid
$K\simeq -1$ the gaps in $\chi''(\vc q,\omega)$ at $\vc q=\Gamma$ and
$\vc q=X$ open gradually from the points of quantum phase transition
from ordered to disordered phase. With much weaker quantum corrections
for ordered phases in the regime of ferromagnetic Kitaev interactions,
the spin liquid is more robust near $K=-1$ as a phase that contains
quantum fluctuations and survives in a broader regime than near $K=1$
when antiferromagnetic Kitaev interactions are disturbed by increasing
(antiferromagnetic or ferromagnetic) Heisenberg interactions.
This behavior is reminiscent of the ferromagnetic Kitaev model in
a weak magnetic field \cite{Tik11}.

\acknowledgments

We thank Giniyat Khaliullin for insightful discussions.
We kindly acknowledge support by Narodowe Centrum Nauki
(NCN, National Science Center) under Project No. 2012/04/A/ST3/00331.
J. R. and J. C. were supported by Czech Science Foundation (GA\v{C}R)
under Project No. GJ15-14523Y and by the project CEITEC 2020 (LQ1601)
with financial support from the Ministry of Education, Youth and Sports
of the Czech Republic under the National Sustainability Programme II.
Access to computing and storage facilities owned by parties and
projects contributing to the National Grid Infrastructure MetaCentrum,
provided under the program ``Projects of Large Research, Development,
and Innovations Infrastructures" (CESNET LM2015042), is acknowledged.
G.~J.~is supported in part by the National Science Foundation under
Grant No. NSF PHY11-25915.

\appendix*

\section{ Comparison between CMFT and linearized CMFT  for a single hexagon}
\label{sec:hex}

Here we compare linearization results for a single hexagon with full
CMFT to see how well linearized CMFT performs as a shortcut method.
It is important to realize that this cluster is not compatible with
stripy or zigzag order because of their four-site magnetic unit cell,
see Fig. \ref{1}(b), and they are suppressed within vast regions of
$\varphi$ compared to the  24-site case. The size of the system allows
for quick CMFT computations and enables detailed comparison between
the two approaches. Moreover, specific problems linked to the above
incompatibility make the $N=6$-site cluster a good test case to
illustrate the linearized CMFT.

\begin{figure}[t!]
\includegraphics[width=8.2cm]{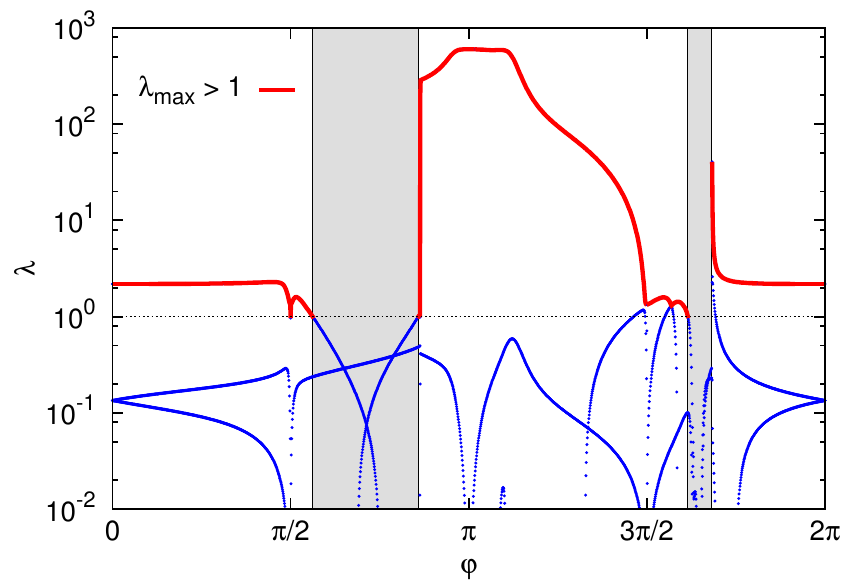}
\caption{
Full linearized CMFT result for a single hexagon. Blue lines represent
all emerged positive eigenvalues $\lambda$, while maximal $\lambda$
larger than 1 is indicated in red.
}
\label{la6}
\end{figure}

\begin{figure}[b!]
\includegraphics[width=8cm]{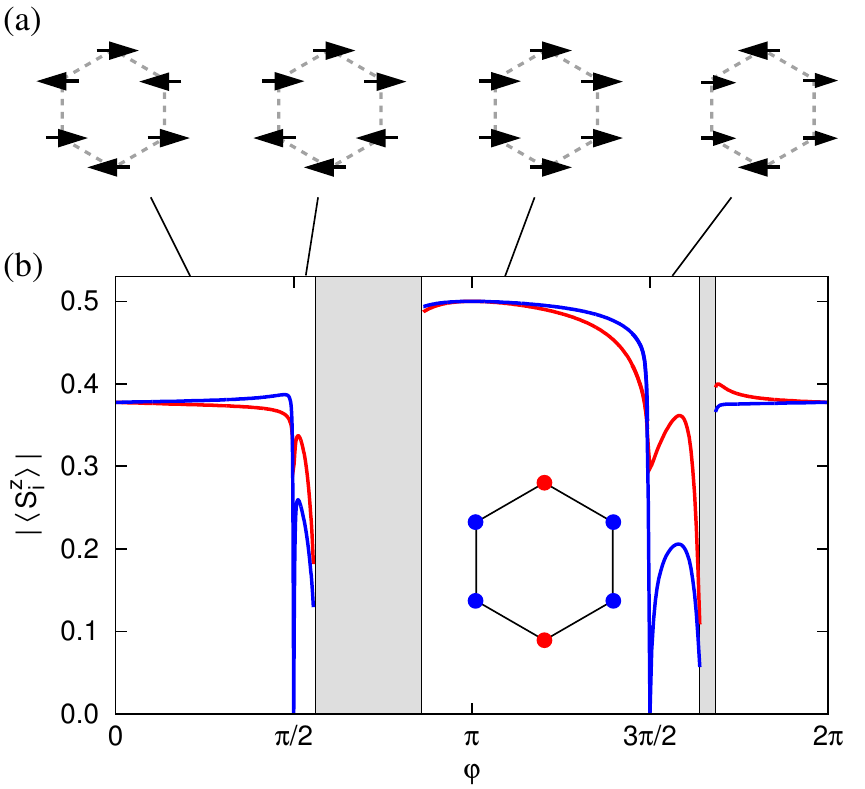}
\caption{
(a) Spin patterns obtained for a single hexagon by CMFT.
From the left: N\'eel, zigzag, FM and stripy.
(b) Phase diagram for a single hexagon determined by
$|\langle S_i^z\rangle|$. Red and blue sites (see inset) are
nonequivalent in the present CMFT due to the approximation given
by Eq. (\ref{SzSz})
which generates the terms $\propto J$ that add to Kitaev term only on
the vertical bonds $\langle ij\rangle\parallel z$ in the MF part of the
Hamiltonian (\ref{ham_in}).
}
\label{sz6}
\end{figure}

Following the procedure described in Sec. \ref{sec:lin}, 6 eigenvalues
$\lambda_i$ are produced for each value of $\varphi$ parameter. The
corresponding spin patterns are inferred by inspecting the eigenvectors.
Only the patterns associated with $\lambda_i>1$ are able to grow during
iterations and eventually stabilize as a self-consistent solution of
full CMFT. Comparison of both methods presented in Figs. \ref{la6} and
\ref{sz6}  provides the phase diagram for a single hexagon:
N\'eel phase for $\varphi\in[0, 0.5)\pi$,
KSL for $\varphi=\frac{\pi}{2}$,
zigzag phase for $\varphi\in(0.5,0.555)\pi$,
disordered region I for $\varphi\in (0.555, 0.864)\pi$,
FM phase for $\varphi\in (0.864,1.5)\pi$,
KSL for $\varphi=\frac{3}{2}\pi$,
stripy phase for $\varphi\in(1.5,1.62)\pi$ (linearization),
$\varphi\in(1.5,1.64)\pi$ (CMFT),
disordered region II for $\varphi\in(1.62,1.684)\pi$ (linearization)
and $\varphi\in(1.64,1.684)\pi$ (CMFT),
and N\'eel phase for $\varphi\in(1.684, 2]\pi$.
In contrast to $N=24$ cluster the two spin-liquid regions
are replaced by single points  $\varphi=\frac{\pi}{2}$ and
$\varphi=\frac{3}{2}\pi$.

Striking difference between phase diagrams for 24-site and 6-site
clusters is the reduction of the zigzag and stripy phases and the
emergence of  two regions of disorder indicated by two gray-shaded
regions. Here all $\lambda_i <1$ and no spin pattern is strong enough
to stabilize. Zigzag pattern emerges from CMFT with
random initial values of $\langle S_i^z\rangle$ without
additional help. Stripy pattern however is more difficult to catch.
As one can see in Fig. \ref{la6}, two different $\lambda_i$ corresponding
to two stripy patterns exchange at $\varphi=1.568\pi$. Unfortunately,
huge parasitic oscillations make these patterns extremely difficult to stabilize
within CMFT. These stem from a large negative $\lambda_i$ that previously
corresponded to FM pattern and decreased rapidly for $\varphi>1.5\pi$.
If one recalls that the equivalent of one iteration in linearized
version of CMFT  is in fact multiplication by $\lambda_{i}$, one can
easily see that large negative $\lambda_i$ would cause oscillations
with an exponentially growing amplitude when performing the iterations
of the self-consistent loop.
To overcome this issue we introduce a damping into a self-consistent
loop by taking
$(1-d)\langle S_i^z\rangle_\mathrm{fin}+d\langle S_i^z\rangle_\mathrm{ini}$
as the new averages. Here $d<1$ is a suitably chosen damping factor.
With this modification CMFT produces one finite stripy order suggested
by linearization. However since the parasitic negative $\lambda_i$
grows enormously in magnitude as we approach the phase boundary an
extreme damping has to be included making the phase boundary hard
to determine by using CMFT.

In conclusion, it is evident that the ordered patterns suggested by
linearization were reproduced by CMFT within regions dictated by the
maximal $\lambda_i>1$. Moreover, the linearized procedure indicated
possible difficulties with stabilizing stripy phases that had to be
cured by a strong damping introduced into the self-consistent loop.


\end{document}